\documentclass[twocolumn,showpacs,preprintnumbers,amsmath,amssymb,prl]{revtex4-1}

\usepackage{amssymb}
\usepackage{graphicx}
\usepackage{bm}

\begin{document}
\title{Strong surface scattering in ultrahigh mobility Bi$_{2}$Se$_{3}$ topological insulator crystals}

\author{N. P. Butch}
\email{nbutch@umd.edu}
\author{K. Kirshenbaum}
\author{P. Syers}

\author{A. B. Sushkov}
\author{G. S. Jenkins}
\author{H. D. Drew}

\author{J. Paglione}
\affiliation{Center for Nanophysics and Advanced Materials, Department of Physics, University of Maryland, College Park, MD 20742}

\date{\today}

\begin{abstract}
While evidence of a topologically nontrivial surface state has been identified in surface-sensitive measurements of Bi$_{2}$Se$_{3}$, a significant experimental concern is that no signatures have been observed in bulk transport. In a search for such states, nominally undoped single crystals of Bi$_{2}$Se$_{3}$ with carrier densities approaching $10^{16}$~cm$^{-3}$ and very high mobilities exceeding 2~m$^2$~V$^{-1}$~s$^{-1}$ have been studied. A comprehensive analysis of Shubnikov de Haas oscillations, Hall effect, and optical reflectivity indicates that the measured electrical transport can be attributed solely to bulk states, even at 50~mK at low Landau level filling factor, and in the quantum limit. The absence of a significant surface contribution to bulk conduction demonstrates that even in very clean samples, the surface mobility is lower than that of the bulk, despite its topological protection.

\end{abstract}

\maketitle
Topological insulators are bulk insulators that feature chiral Dirac cones on their surface. This surface state, which is protected by time-reversal symmetry, is of fundamental interest and may have potential for spintronics and quantum computation applications. A small group of semimetallic stoichiometric chalcogenides have been theoretically shown to have the properties of three-dimensional topological insulators: Bi$_2$Se$_3$, Sb$_2$Te$_3$, and Bi$_2$Te$_3$~\cite{Xia09,Zhang09}. The chiral surface state has been investigated by surface probes, notably angle-resolved photoemission spectroscopy (ARPES) \cite{Xia09,Chen09,HsiehNat09} and scanning tunneling microscopy (STM) \cite{ZhangSTM09,Alpichshev09} and similar studies exist of the alloy Bi$_{1-x}$Sb$_x$ \cite{HsiehSci09,Roushan09}. The observation of chiral Dirac cones and forbidden backscattering in these experiments has generated a great deal of excitement.

In principle, electrical conduction in topological insulators should occur only at the surface, but in practice these stoichiometric materials have been known for decades to be low carrier density metals. Bi$_2$Se$_3$ is expected to have a 300~meV direct bandgap at zone center \cite{Xia09,Zhang09}, yet is almost always \emph{n}-type, long believed the result of charged Se vacancies. It remains an open question whether a bulk insulating state can be achieved in stoichiometric undoped Bi$_2$Se$_3$. This problem has been recently sidestepped by counter-doping with Ca \cite{Xia09,Hor09,Checkelsky09}, although this procedure introduces further defects. Over a range of carrier densities, ARPES data show one conduction band at zone center \cite{Xia09,HsiehNat09}, and bulk Shubnikov de Haas measurements indicate that the Fermi surface is ellipsoidal \cite{Kulbachinskii99}.

There have been several recent studies of bulk transport in Bi$_2$Se$_3$. A thorough study of angle dependent Shubnikov de Haas (SdH) oscillations on a sample with a moderate carrier density has mapped the anisotropy of the Fermi surface \cite{Eto10}. A comparative investigation demonstrated that while ARPES may identify a Fermi level in the band gap, SdH oscillations in similar samples always show metallic behavior, suggesting that the Fermi level may vary between bulk and surface in Bi$_2$Se$_3$ \cite{Analytis10}. This raises questions about the correspondence between surface and bulk studies on the same material. While it is possible to tune the Fermi level of Bi$_2$Se$_3$ into the gap via Ca doping, the resulting samples show unusual magnetoresistance fluctuations, but no SdH oscillations from the surface state \cite{Checkelsky09}. In contrast, field-dependent oscillations ascribed to the surface state have been observed in tunneling spectra measured on Bi$_2$Se$_3$ thin films \cite{Cheng10,Hanaguri10} and Aharonov-Bohm interferometry of the surface state is reported in Bi$_2$Se$_3$ nanoribbons \cite{Peng10}. Because SdH oscillations attributable to surface states were identified in intrinsically disordered Bi$_{1-x}$Sb$_{x}$ alloys \cite{Taskin10}, it is intriguing that analogous bulk transport phenomena have not been seen in stoichiometric Bi$_2$Se$_3$.

In order to look for signatures of the novel surface state in bulk nominally undoped Bi$_2$Se$_3$, single crystals were synthesized over a wide range of carrier concentrations. Their longitudinal and transverse electrical resistivity were systematically characterized as a function of field and temperature, and infrared reflection and transmission were studied. This investigation confirms that all measured transport properties can be ascribed to bulk electron-like carriers, despite the samples having the lowest reported bulk carrier densities and highest mobilities. Even at low Landau level filling factor and in the quantum limit, no transport signature of surface states is apparent, placing strong constraints on the mobility of surface carriers in these high quality samples.

Single crystals of Bi$_2$Se$_3$ were prepared by melting high purity bismuth (6N) and selenium (5N) in sealed quartz ampoules. Multiple batches were prepared with varying bismuth/selenium ratios and heating conditions, which were responsible for variations in the carrier concentrations of the samples. Similar trends were reported decades ago \cite{Hyde74} and more recently \cite{Analytis10}. The resultant crystals cleaved easily perpendicular to the $c_3$ trigonal axis. Four-probe measurements of longitudinal and transverse electrical transport between 1.8~K and 300~K were conducted using a Quantum Design 14~T PPMS.  The current flowed in the plane perpendicular to $c_3$ in both longitudinal and transverse geometries, and the magnetic field $H$ was always applied perpendicular to applied current. Electrical transport measurements at lower temperatures were performed in a dilution refrigerator with a rotating sample stage and 15/17~T magnet. Optical measurements were performed using a Bomem DA3 FTIR spectrometer. Carrier density $n$ was determined by measurements of Hall effect, SdH oscillations, and fits to optical reflectivity data, which always yielded identical results within experimental uncertainty.

\begin{figure}
    {\includegraphics[width=3.4in]{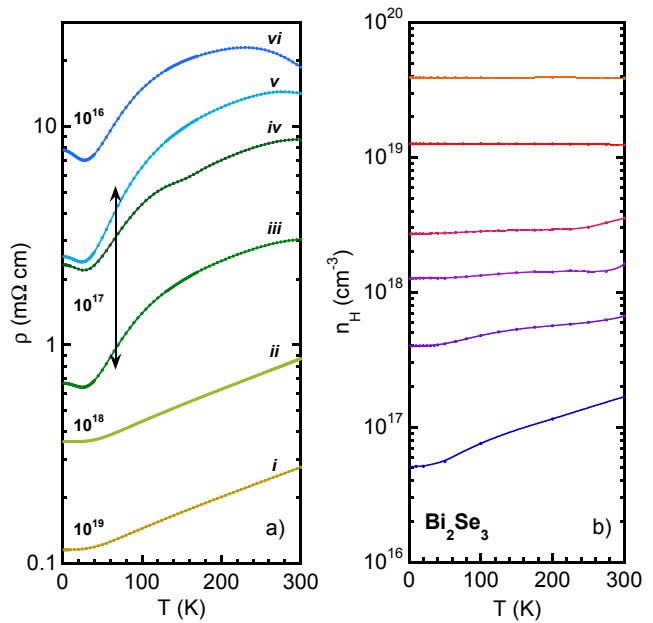}}
    \caption{(Color online) a) Comparison of the electrical resistivity $\rho(T)$ between samples of Bi$_2$Se$_3$. Carrier densities $n$ (cm$^{-3}$) are estimated from SdH oscillations: \emph{\textbf{i}}~$1 \times 10^{19}$, \emph{\textbf{ii}}~$5.3 \times 10^{18}$, \emph{\textbf{iii}}~$4.9 \times 10^{17}$, \emph{\textbf{iv}}~$3.7 \times 10^{17}$, \emph{\textbf{v}}~$3.3 \times 10^{17}$, \emph{\textbf{vi}}~$ \sim 10^{16}$. In samples with $n < 10^{18}$~cm$^{-3}$, a shallow minimum develops at 30~K. b) Comparison of Hall carrier density $n_\mathrm{H}$ between various samples. Low temperature values span 3 orders of magnitude.}
    \label{batchcompare}
\end{figure}

The temperature dependence of the longitudinal electrical resistivity $\rho(T)$ is summarized in Figure~\ref{batchcompare}a. The values of $n$ in these samples were estimated via analysis of the SdH oscillations in $\rho(H)$ and approximating the Fermi surface as spherical, which is sufficient for purposes of comparison at low $n$ \cite{Kulbachinskii99}. In sample \emph{\textbf{vi}}, no SdH oscillations are observed up to 14~T, and the estimated $n = 5 \times 10^{16}$~cm$^{-3}$ is based on Hall measurements on samples from the same batch (Fig.~\ref{batchcompare}b). For $n>10^{18}$~cm$^{-3}$, the $\rho(T)$ data reflect good metallic conductivity and generally, as $n$ decreases, the samples become more resistive. For $n < 10^{17}$~cm$^{-3}$, a negative slope develops above 250~K, which reflects a crossover to activated behavior at $T>300$~K \cite{Kohler75}. For $n<10^{18}$~cm$^{-3}$, the $\rho(T)$ data also develop a shallow local minimum at 30~K, which is associated with a relatively small upturn that saturates as $T \rightarrow 0$.

\begin{figure}
    {\includegraphics[width=3.4in]{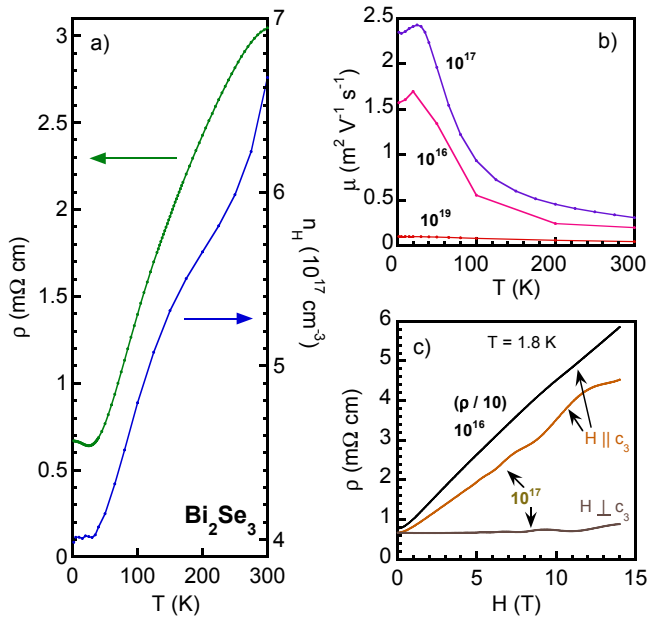}}
    \caption{(Color online) a) A comparison of electrical resistivity $\rho(T)$ (\emph{\textbf{iii}} from Fig.~\ref{batchcompare}) and Hall carrier density $n_\mathrm{H}(T)$ in two samples from the same batch. The minimum in $\rho(T)$ coincides with the saturation of $n_\mathrm{H}(T)$ below 30~K. Despite increasing $n_\mathrm{H}$, $\rho$ increases with $T$ up to 300~K. b) The calculated mobility exceeds 2~m$^2$~V$^{-1}$~s$^{-1}$ at low $T$ (labeled $10^{17}$). The mobilities of samples with lower and higher $n$ are compared. c) Field orientation dependence of $\rho$ at 1.8~K. Compare to $\rho(H)$ of sample \emph{\textbf{vi}} (rescaled by factor of 10), which exhibits no SdH oscillations. }
    \label{mobility}
\end{figure}

In Figure~\ref{batchcompare}b, the $T$ dependence of the Hall carrier densities $n_\mathrm{H} = (R_\mathrm{H} e)^{-1}$ is shown for a range of samples. Values of the Hall coefficient $R_\mathrm{H}$ were determined via linear fits to symmetrized transverse magnetoresistance data, which are linear to 5~T over the entire $T$ range, except for at lowest $n$ where linearity extends to 2~T at low $T$. Generally, $n_\mathrm{H}(T)$ has a gentle $T$-dependence, consistent with a gradual crossover from extrinsic to intrinsic conduction ($E_\mathrm{g}>100$~meV) \cite{Kohler75}. However, metallic $\rho(T)$ (Fig.~\ref{batchcompare}a) is always observed, indicating that phonon scattering dominates any increase in conductivity due to increasing carrier number for most of the $T$ range. A detailed comparison of $\rho(T)$ (sample \emph{\textbf{iii}}) and $n_\mathrm{H}(T)$ between two samples from the same batch is shown in Fig.~\ref{mobility}a. These data highlight two regimes: below 30~K, where $n_\mathrm{H}$ is constant and $\rho(T)$ exhibits an upturn, and between 150 and 250~K, where there is a change in curvature in $n_\mathrm{H}(T)$ and $\rho(T)$. The magnitude of these latter features is greatest in samples with $n < 10^{17}$~cm$^{-3}$. The two $T$ ranges are accentuated in a plot of the Hall mobility $\mu(T) = R_\mathrm{H}(T)/\rho(T)$ (Fig.~\ref{mobility}b), highlighting the substantial low-$T$ mobility $\mu>2$~m$^2$~V$^{-1}$~s$^{-1}$ in these samples, which is the highest value reported for Bi$_2$Se$_3$ and corresponds to a mean free path longer than 300~nm. A comparison to $\mu(T)$ of batches with $n \sim 10^{16}$~cm$^{-3}$ and $n \sim 10^{19}$~cm$^{-3}$ indicates that $\mu$ does not simply increase with decreasing $n$, although at low $n$ it still exceeds $1$~m$^2$~V$^{-1}$~s$^{-1}$.

\begin{figure}
    {\includegraphics[width=3.4in]{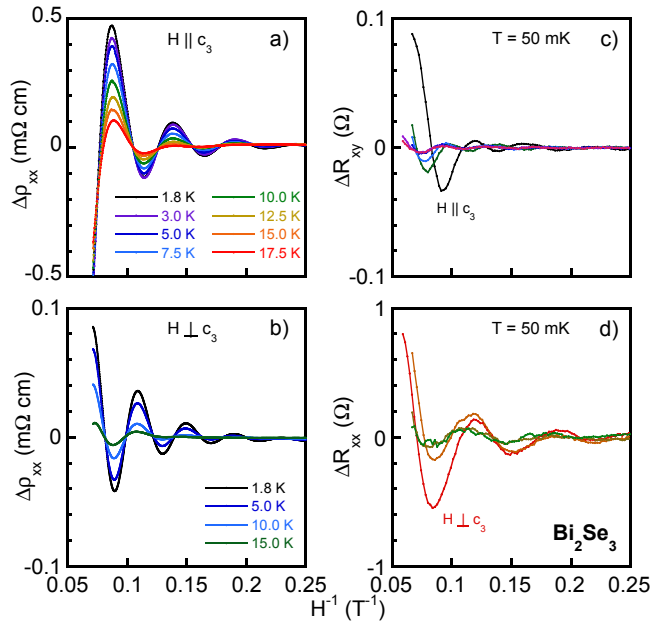}}
    \caption{(Color online) Temperature dependence of SdH oscillations with a) $H \parallel c_3$ trigonal axis and b) $H \perp c_3$. Oscillation frequencies $f_\mathrm{SdH}=20$~T and 25~T, respectively, and the mass $m_{SdH} \approx 0.15 m_\mathrm{e}$. SdH oscillations from two different samples at 50~mK in the c) transverse and d) longitudinal MR. The largest amplitude and smallest frequency in c) is observed with $H \parallel c_3$ and the frequency ranges from 20~T to 27~T. In d) the largest amplitude is observed with $H \perp c_3$ and $f_\mathrm{SdH} = 15$~T. Only one SdH frequency is observed in each measurement.}
    \label{SdH}
\end{figure}

In Fig.~\ref{mobility}c, $\rho(H)$ of sample \emph{\textbf{iii}} is shown. In addition to the dramatic anisotropy, oscillations are readily apparent. The non-oscillatory part is sufficiently well described by quadratic polynomial or weak power law functions. In contrast, sample \emph{\textbf{vi}} exhibits no oscillations. The $T$-dependence of SdH oscillations is shown in Fig.~\ref{SdH} in panels a) and b) for two different field orientations (c.f. Fig.~\ref{mobility}c). The SdH oscillations have a frequency $f_\mathrm{SdH}=20$~T when $H \parallel c_3$ and $f_\mathrm{SdH}=25$~T when $H \perp c_3$, which correspond to an ellipsoidal Fermi surface with $n = 6.3 \times 10^{17}$~cm$^{-3}$. A Lifshitz-Kosevich fit to the $T$-dependence of oscillation amplitude yields masses $m_\parallel = 0.14(1) m_\mathrm{e}$ and $m_\perp = 0.16(1) m_\mathrm{e}$, where $m_e$ is the bare electron mass, consistent with other reports \cite{Analytis10,Eto10}. In order to check whether lowering the temperature would uncover a second frequency of SdH oscillations arising from the surface state, measurements were performed at 50~mK on two samples with $n_\mathrm{H} = 4-5 \times 10^{17}$~cm$^{-3}$ from the same batch as sample \emph{\textbf{iv}}. The field angle dependence of both longitudinal and transverse resistance are shown in Fig.~\ref{SdH} panels c) and d). Figure~\ref{SdH}c shows oscillations in the transverse magnetoresistance $R_{xy}(H)$, whose frequencies range from 20~T with $H \parallel c_3$ to 27~T with $H \perp c_3$ ($n = 7 \times 10^{17}$~cm$^{-3}$). As expected, the magnitude of the effective Hall signal diminishes as $H$ rotates into the plane and the evolution of the oscillation frequency reflects the expected Fermi surface anisotropy. The longitudinal SdH oscillations are shown in Fig.~\ref{SdH}d. The most prominent oscillations, with $f_\mathrm{SdH} = 15$~T ($n = 3 \times 10^{17}$~cm$^{-3}$) occur with $H \perp c_3$. In this field orientation, the electrons are almost all in the first Landau level. Weak oscillations with $H \parallel c_3$ have $f_\mathrm{SdH} = 11$~T, yielding an anisotropy close to that of the other two samples. The lack of bulk SdH oscillations in sample \emph{\textbf{vi}} (Fig.~\ref{mobility}c) having even lower $n$ is due to its low estimated $f_\mathrm{SdH} = 4$~T, meaning that most of the accessible field range is in the bulk quantum limit.

\begin{figure}
\includegraphics[width=\columnwidth]{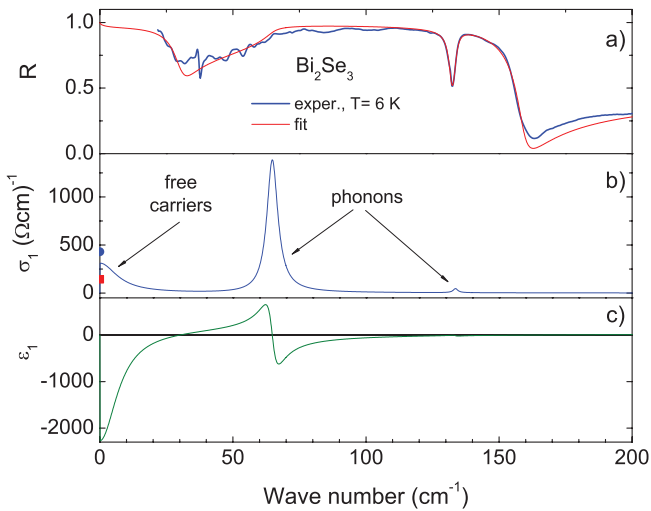}
    \caption{(Color online) Spectra of a) reflectivity, b) optical conductivity, and c) dielectric constant. Curves b) and c) were calculated from the fit to the reflectivity data in a). For comparison, the blue circle and red square in b) are dc values from sample \emph{\textbf{iv}} for 6~K and room temperature, respectively.}
\label{optics}
\end{figure}

Samples were also characterized by infrared reflection and transmission, which confirm the estimates of $n$ based on Hall and SdH data. Figure~\ref{optics}a shows an experimental reflectivity spectrum of a sample from the same batch as sample \emph{\textbf{iv}}. Data were taken at 6~K and are presented with a fit to a sum of Lorentz oscillators to model the complex dielectric function.  The fit shown in Fig.~\ref{optics}a includes only bulk carriers as a Drude-type term together with two low frequency phonons. A strong low frequency phonon is centered at 67~cm$^{-1}$ with a plasma frequency of 640~cm$^{-1}$ and relaxation rate of 5~cm$^{-1}$ and a second weaker phonon is centered second at 124~cm$^{-1}$ with a plasma frequency of 76~cm$^{-1}$ and a width of 2~cm$^{-1}$.  The calculated optical conductivity is shown in Fig.~\ref{optics}b together with two dc values (sample \emph{\textbf{iv}}) for rough comparison. This yields a bare plasma frequency of free carriers to be 382~cm$^{-1}$, corresponding to $n=2.4\times10^{17}$~cm$^{-3}$ for $0.15 m_\mathrm{e}$. The effective scattering rate $\gamma = 8$~cm$^{-1}$ ($\tau = 0.8$~ps) is unusually low for bulk carriers, consistent with a high mobility sample. Figure~\ref{optics}c shows the real part of the dielectric function. The fit parameters are in agreement with those from the transmission spectra of very thin crystals. An important difference, however, is that in transmission, a very broad Drude surface contribution ($\gamma = 100$~cm$^{-1}$) is detected.

For small $n$, the competition between negative $\varepsilon_1$ of the free carriers and the positive $\varepsilon_1$ from the phonons gives rise to an additional low frequency plasma edge ($\varepsilon_1 = 0$) that leads to a transmission window near 30~cm$^{-1}$. The strong bismuth-dominated low frequency phonon produces a large static dielectric constant $\varepsilon_1(0) \approx 100$ for Bi$_2$Se$_3$. This implies a large reduction in the Coulomb potential and thus the scattering rate from ionized impurities, which can account for the small scattering rate of the bulk carriers and the good metallic behavior of this doped semiconducting material, and protecting against localization at low $n$. The Coulomb mediated electron-electron interaction is also reduced \cite{Baldwin80} and, along with the presence of low frequency optical phonons, implies a dominance of the electron-phonon interaction both for the carrier mobility and for the electron-electron interaction.

ARPES measurements show that surface states in Bi$_2$Se$_3$ always have a larger Fermi surface than the bulk and that the areal difference increases as the chemical potential decreases \cite{HsiehNat09}. Quantum oscillations arising from the surface should have an easily identifiable higher frequency of about 300~T ($k_\mathrm{F}=0.1$~\AA$^{-1}$) when the chemical potential is near the bottom of the conduction band. However, no higher frequency oscillation is ever observed in our measurements, even in the quantum limit, where no bulk SdH oscillations exist to obscure surface signal. As the existence of the surface states is not in doubt, the absence of their transport signature can only arise from very high scattering rates. A conservative assumption that the chemical potential is the same at the surface and bulk, consistent with ARPES surface aging data \cite{HsiehNat09}, yields $n_\mathrm{surf} \approx 10^{13}$~cm$^{-2}$. In our samples, the bulk areal carrier density is about 100 times that of the surface, while the peak-to-noise ratio in the Fourier transform of the SdH oscillations is greater than 500, so the surface mobility has to be at least 5 times worse than the bulk mobility for SdH oscillations to be below our detection threshold. We calculate scattering times $\tau_\mathrm{bulk}=2$~ps and $\tau_\mathrm{surf}<0.4$~ps. The latter value is generous compared to an upper limit of $\tau_\mathrm{surf}<0.05$~ps determined from our transmission measurements. These estimates indeed indicate substantial surface scattering, yet are entirely consistent with ARPES data, which show significant surface band broadening, with $\tau=0.02$~ps \cite{Xia09} or 0.04~ps \cite{Park10} even for samples cleaved in high vacuum.

It is clear that topological protection does not guarantee high surface mobility, so increasing the surface conduction will require considerable effort. Reducing sample thickness is one possibility, but bulk transport properties appear to be insensitive to sample thickness down to 5~$\mu$m \cite{Analytis10}. Even in 50~nm thick nanoribbons, the observed $f_\mathrm{SdH} =  70$~T and 110~T \cite{Peng10} are well below 300~T and are not likely due to surface states. Relative surface conduction may alternatively be improved by further reducing bulk $n$. Attempts to do so via chemical doping yield no surface SdH oscillations \cite{Checkelsky09}, possibly due to even further decreased mobility. On the other hand, doping may improve the surface signal if it preferentially decreases the bulk mobility without reducing surface mobility, which may be possible considering the already high surface scattering rate. A cleaner approach is to instead preferentially improve the surface mobility via controlled surface preparation. Such a task will be challenging, as it is difficult to improve upon the vacuum cleaving employed in ARPES studies, which yields broad surface state linewidths. Nonetheless, high surface mobility is a prerequisite for precision experiments like optical Kerr rotation and future applications potential of topological insulators.

We note a recent report of the observation of surface SdH oscillations in pulsed fields \cite{Analytis10a}. The cleaved surfaces of the Sb-doped Bi$_2$Se$_3$ were minimally exposed, yet the requirement of high magnetic fields points to strong surface scattering even under controlled conditions.

\begin{acknowledgments}
We thank M. S. Fuhrer, S. K. Goh, R. M. Lutchyn, and M. L. Sutherland for helpful discussions. This work was supported in part by the National Science Foundation MRSEC under Grant No. DMR-0520471.  NPB is supported by CNAM.
\end{acknowledgments}


\begin{thebibliography}{99}

\bibitem{Zhang09} H. Zhang, \emph{et al}., Nat. Phys. \textbf{5}, 438 (2009).

\bibitem{Xia09} Y. Xia, \emph{et al}., Nat. Phys. \textbf{5}, 398 (2009). 

\bibitem{Chen09} Y. L. Chen, \emph{et al}., Science \textbf{325}, 178 (2009). 

\bibitem{HsiehNat09} D. Hsieh, \emph{et al}., Nature (London) \textbf{460}, 1101 (2009). 

\bibitem{ZhangSTM09} T. Zhang, \emph{et al}., Phys. Rev. Lett. \textbf{103}, 266803 (2009).

\bibitem{Alpichshev09} Z. Alpichshev, \emph{et al}., arXiv:0908.0371v2 (unpublished).

\bibitem{HsiehSci09} D. Hsieh, \emph{et al}., Science \textbf{323}, 919 (2009). 

\bibitem{Roushan09} P. Roushan, \emph{et al}., Nature (London) \textbf{460}, 1106 (2009). 

\bibitem{Hor09} Y. S. Hor, \emph{et al}., Phys. Rev. B \textbf{79}, 195208 (2009).

\bibitem{Checkelsky09} J. G. Checkelsky, \emph{et al}., Phys. Rev. Lett. \textbf{103}, 246601 (2009).

\bibitem{Kulbachinskii99} V. A. Kulbachinskii, \emph{et al}., Phys. Rev. B \textbf{59}, 15733 (1999).

\bibitem{Eto10} K. Eto, Z. Ren, A. A. Taskin, K. Segawa, and Y. Ando, Phys. Rev. B \textbf{81}, 195309 (2010).

\bibitem{Analytis10} J. G. Analytis, \emph{et al}., Phys. Rev. B \textbf{81}, 205407 (2010).

\bibitem{Cheng10} P. Cheng, \emph{et al}., arXiv:1001.3220 (unpublished).

\bibitem{Hanaguri10} T. Hanaguri, K. Igarashi, M. Kawamura, H. Takagi, and T. Sasagawa, arXiv:1003.0100v1 (unpublished).

\bibitem{Peng10} H. Peng, \emph{et al}., Nat. Mat. \textbf{9}, 225 (2010).

\bibitem{Taskin10} A. A. Taskin, K. Segawa, and Yoichi Ando, arXiv:1001.1607v1 (unpublished).

\bibitem{Hyde74} G. R. Hyde, H. A. Beale, I. L. Spain, and J. A. Woollam, J. Phys. Chem. Solids \textbf{35}, 1719 (1974).

\bibitem{Kohler75} H. K\"{o}hler and A. Fabricius, Phys. Stat. Sol. \textbf{b71}, 487 (1975).

\bibitem{Baldwin80} S. Baldwin and H. D. Drew, Phys. Rev. Lett. \textbf{45}, 2063 (1980).

\bibitem{Park10} S. R. Park, \emph{et al}., Phys. Rev. B \textbf{81}, 041405(R) (2010).

\bibitem{Analytis10a} J. G. Analytis, \emph{et al}., arXiv:1003.1713v1 (unpublished).


\end{thebibliography}
\end{document}